\documentclass[11pt]{article}
\pdfoutput=1
\usepackage{jheppub}
\usepackage[usenames,dvipsnames,table]{xcolor}
\usepackage{graphicx,amsmath,amssymb,amsthm,multirow,array,bm,bbm,esint}
\usepackage[mathscr]{eucal}
\usepackage{subfig}

\newcommand\beq{\begin{equation}}
\newcommand\eeq{\end{equation}}
\def\be{\begin{eqnarray}}
\def\ee{\end{eqnarray}}

\def\Dslash{\,\,{\raise.15ex\hbox{/}\mkern-12mu D}}
\def\Dbarslash{\,\,{\raise.15ex\hbox{/}\mkern-12mu {\bar D}}}
\def\delslash{\,\,{\raise.15ex\hbox{/}\mkern-9mu \partial}}
\def\delbarslash{\,\,{\raise.15ex\hbox{/}\mkern-9mu {\bar\partial}}}
\def\pslash{\,\,{\raise.15ex\hbox{/}\mkern-9mu p}}
\def\calDslash{\,\,{\raise.15ex\hbox{/}\mkern-12mu {\cal D}}}

\newcommand{\sign}{{\rm sign}}

\def\lae{\mathrel{\mathop{\smash{\lower .5 ex \hbox{$\stackrel<\sim$}}}}}
\def\lae{\mathrel{\mathop{\smash{\lower .5 ex \hbox{$\stackrel>\sim$}}}}}

\title{Bosonizing three-dimensional quiver gauge theories }

\author[a]{Kristan Jensen}
\author[b]{ and Andreas Karch}
\affiliation[a]{Department of Physics and Astronomy, San Francisco State University, San Francisco, CA 94132}
\affiliation[b]{Department of Physics, University of Washington, Seattle, WA 98195-1560}

\preprint{\today}

\emailAdd{kristanj@sfsu.edu,akarch@uw.edu}

\abstract{We start with the recently conjectured 3d bosonization dualities and gauge global symmetries to generate an infinite sequence of new dualities. These equate theories with non-Abelian product gauge groups and bifundamental matter. We uncover examples of Bose/Bose and Fermi/Fermi dualities, as well as a sequence of dualities between theories with scalar matter in two-index representations. Our conjectures are consistent with level/rank duality in massive phases.}

\begin{document}
\maketitle

\section{Introduction}

Chern-Simons (CS) gauge theories in three dimensions enjoy ``level/rank'' dualities. The most basic of these relates
\beq
U(N)_{-k} \quad\leftrightarrow \quad SU(k)_N\,,
\eeq
for $N,k\geq 0$, where the subscript indicates the Chern-Simons level. The two theories are dual in that the observables in both theories have rigorously been shown to be identical~\cite{Naculich:1990pa,Mlawer:1990uv,Nakanishi:1990hj}.

There has been a growing body of evidence for a large family of non-supersymmetric dualities between Chern-Simons theories coupled to matter (see e.g.~\cite{Aharony:2011jz,Giombi:2011kc,Aharony:2012nh,Aharony:2015mjs,Hsin:2016blu}). These may be thought of as a generalization of level/rank dualities, where we add flavor fields to both sides of the duality and tune to a conformal field theory (CFT).
Since they typically relate a theory with fermions to a theory with bosons, these dualities are sometimes called ``3d bosonization''. The two basic pairs that will play a major role for us were precisely formulated by Aharony \cite{Aharony:2015mjs} and read:
\begin{eqnarray}
\label{base1}
SU(N)_{-k+N_f/2} \mbox{ with } N_f \mbox{ Dirac fermions } &\quad \leftrightarrow \quad&
U(k)_N \mbox{ with } N_f \mbox{ scalars, } \\
 \label{base2} U(N)_{-k+N_f/2} \mbox{ with } N_f \mbox{ Dirac fermions } &\quad \leftrightarrow \quad&
SU(k)_N \mbox{ with } N_f \mbox{ scalars. }
\end{eqnarray}
On the fermion side of the story, all interactions are due to the Chern Simons gauge field. On the scalar side, there is a quartic potential for the scalar field, with the quartic coupling set to its non-trivial Wilson-Fisher (WF) fixed point value, that is, to the attractive IR fixed point. Both dualities require $N_f \leq k$.

The dualities~\eqref{base1} and~\eqref{base2} are unproven. The evidence for them largely comes from computations at large $N$ and large $k$ with $N/k$ finite~\cite{Aharony:2011jz,Giombi:2011kc,Aharony:2012nh,Aharony:2012ns}, the matching of global symmetries and their `t Hooft anomalies~\cite{Benini:2017dus}, and consistency with level/rank duality. What is meant by this last item is the following. On both sides of~\eqref{base1} and~\eqref{base2}, we can deform by a mass operator to obtain a topological field theory (TFT) at low energies, with different TFTs arising depending on the sign choice for the deformation. The original CFT can be thought of at the critical point describing the phase transition between these distinct topological phases. The conjectured dualities are consistent with level/rank duality in that the TFTs in the massive phases of one side are level/rank dual to those on the other. For example, $SU(N)_{-k+N_f/2}$ with $N_f$ fermions and a negative mass flows to $SU(N)_{-k}$ in the infrared, which is level/rank dual to the $U(k)_N$ one obtains by deforming $U(k)_N$ with $N_f$ scalars by a positive mass-squared.

For the special case of $N=k=N_f=1$ the conjectures~\eqref{base1} and~\eqref{base2} can be treated as ``seed dualities" and used to derive a web \cite{Karch:2016sxi,Seiberg:2016gmd} (with closely related work in \cite{Murugan:2016zal}) of Abelian dualities, many of which have been independently conjectured and play an important role in recent discussions of the quantum Hall effect at half filling \cite{Son:2015xqa,ws,mv,mam}. When $N_f=1$, both sides of the dualities have a $U(1)$ global symmetry. An important strategy to derive new dualities from old was to couple this global symmetry to a background gauge field, which was then promoted to a new dynamical field. This procedure was particularly interesting as the $U(1)$ symmetry couples very differently on both sides of the duality. The $SU(N)$ theories have a conserved ``particle number'' or  ``baryon number.'' The matter fields carry charge under this symmetry and correspondingly the background gauge field couples directly to the matter fields via the usual minimal coupling in the kinetic terms. On the other hand, in the $U(N)$ gauge theories, there is a $U(1)$ baryon number, but it is part of the gauge group. All physical observables are neutral under it and it does not correspond to a physical global symmetry. However, the presence of a dynamical $U(1)$ gauge field $a_{\mu}$ leads to a identically conserved topological current $j^{\mu} = \frac{1}{2\pi}\epsilon^{\mu \nu \rho} \partial_{\nu} a_{\rho}$. The conserved charge that comes from this current is monopole number, and the duality equates particle number on one side to monopole number on the other. The background gauge field $A_{\mu}$ couples to this monopole current via a BF coupling, that is through a term in the Lagrangian proportional to $\epsilon^{\mu \nu \rho} A_{\mu} \partial_{\nu} a_{\rho}$.

In this note we apply the same strategy away from $N=k=N_f=1$. We treat the conjectures~\eqref{base1} and~\eqref{base2} as seed, or base, dualities, couple global symmetries on both sides to background fields, and promote the latter to dynamical gauge fields. Assuming that the theories so obtained can be tuned to non-trivial CFTs, we thereby derive novel dualities based on products of non-Abelian groups with bifundamental matter, that is matter charged under two different gauge groups. These type of gauge theories are often known as ``quivers.'' This leads to an infinite sequence of new dualities for quiver gauge theories, and in this note we only focus on some examples that showcase the sort of dualities one gets this way.

Our dualities can be summarized by a simple rule. Given a quiver theory $Q$, we can obtain a dual quiver $Q'$ in which we dualize a single node in $Q$. For example, given a $U(k)_N$ node in $Q$ coupled to bosons, we replace it with a $SU(N)_{-k+N_f/2}$ node coupled to fermions to obtain $Q'$.

In Chern-Simons-matter theories, the gauge field fluctuations are infinitely massive and the only dynamical degrees of freedom are the matter fields. So even though formally the theories in \eqref{base1} and \eqref{base2} describe both matrix- and vector-valued fields, for practical purposes the theories are vector-like. Correspondingly the large $N$, finite $N_f$ limit of these models is soluble, and, in fact, the best evidence for the dualities is in this limit. Once we promote the duality to product groups with bifundamental matter, not even this limit is easily tractable due to the dynamics of the $SU(N_f)$ gauge fields. In the limit where $N_f$ is of order $N$, we are describing gauge theories which have a genuine matrix large $N$ limit.

Besides a richer large $N$ limit, our program leads to qualitatively new bosonization dualities. For example, we find a non-Abelian Bose/Bose duality which, for particular choices of ranks and levels, can be orbifolded to obtain a new duality between $U(k)_N$ theory with a symmetric two-index scalar and $SO(2N)_{-k}$ theory with an adjoint scalar. This last duality is the first example of a sequence of non-supersymmetric dualities we know of with matter fields appearing in two-index representations of a gauge group.

We are unable to prove our new conjectures, but we do perform some basic consistency checks. Because our starting point is to take seed dualities whose global symmetries are already known to match, the global symmetries of our new dualities also match.\footnote{Above, we mentioned that one of the pieces of evidence for the dualities~\eqref{base1} and~\eqref{base2} is that the global symmetries and their `t Hooft anomalies match~\cite{Benini:2017dus}. The alert reader ought to worry whether or not it is legal to then gauge flavor symmetries. The `t Hooft anomalies in question arise only because a quotient of the naive global symmetry acts faithfully on both sides of~\eqref{base1} and~\eqref{base2}, which alters the quantization of the levels of flavor Chern-Simons terms. For generic choices of $k,N,$ and $N_f$, those levels have a fractional part. When this is the case, the theories on both sides cannot be given a purely three-dimensional definition in a general flavor background. This is an obstruction to gauging the flavor symmetry. However there is a simple way out: we simply add a decoupled discrete gauge sector, or equivalently, we gauge the naive global symmetry rather than the quotient that acts faithfully.} A more refined check is to verify that our conjectures are consistent with level/rank duality. As with the seed dualities~\eqref{base1} and~\eqref{base2}, we find an exact map of the various topological phases that can be obtained after deforming both sides by mass operators. This gives some hope that our new conjectures are on the right track.

Supersymmetry may give another test. There has been promising recent work (see e.g.~\cite{Aharony:2012ns,Jain:2013gza,Gur-Ari:2015pca,Kachru:2016rui,Kachru:2016aon}) aimed at deriving the seed dualities~\eqref{base1} and~\eqref{base2} from supersymmetric (SUSY) dualities, both Giveon/Kutasov duality~\cite{Giveon:2008zn} and mirror symmetry~\cite{Intriligator:1996ex}. The idea is to take an $\mathcal{N}=2$ dual pair, turn on SUSY-breaking relevant deformations, and find the non-SUSY bosonization dualities in the IR. This approach has been particularly successful at large $N$ and large $k$~\cite{Jain:2013gza,Gur-Ari:2015pca} where both sides of the SUSY and non-SUSY dualities are under control. It would be very interesting if our new quiver dualities can also be obtained as the offspring of a SUSY duality, although since our theories are matrix-like, we fear it may not be possible to test this in a controlled large $N$ limit.

\section{A simple example}
\subsection{Basic duality} \label{basicduality}

Let us demonstrate the basic idea with a simple example. We conjecture a duality between a product gauge group with a bifundamental scalar $X$ and a different product gauge group with a bifundamental fermion $\psi$ as summarized in the following table.
\vskip10pt
\begin{center}\hfill
\begin{minipage}[h]{0.27\hsize} \centering
\begin{tabular}{|c|c|c|}
\hline
\multicolumn{3}{|c|}{\bf Theory A} \\
\hline
\hline
&$U(k_1)_{N_1}$ & $U(k_2)_{N_2}$ \\
\hline
$X$ & $\Box$ & $\Box$ \\
\hline
\end{tabular}
\end{minipage}
\quad {\huge $\leftrightarrow$} \quad
\begin{minipage}[h]{0.32\hsize}\centering
\begin{tabular}{|c|c|c|}
\hline
\multicolumn{3}{|c|}{\bf Theory B} \\
\hline
\hline
&$SU(N_1)_{-k_1 + \frac{k_2}{2}}$ & $U(k_2)_{N_2+ \frac{N_1}{2}}$ \\
\hline
$\psi$ & $\Box$ & $\Box$ \\
\hline
\end{tabular}
\end{minipage}
\hfill
\begin{minipage}[h]{0.1\hsize} \beq \label{dualone} \eeq \end{minipage}
\end{center}
\vskip10pt
The boxes in the second line indicate that $X$ and $\psi$ transform in the fundamental representation of the gauge group factors. $X$ is a WF scalar, that is, Theory A has a scalar potential $V= m^2|X|^2 + \lambda |X|^4$ with $m^2$ and $\lambda$ tuned to a fixed point, assuming that it exists. The duality is valid for $k_1 \geq k_2$, which we can chose without loss of generality. If $k_2 > k_1$ we can simply exchange the two gauge group factors.

This duality can be derived from the seed~\eqref{base1} by taking $k=k_1$, $N=N_1$, and $N_f=k_2$, tuning the flavor Chern-Simons level on both sides to match, and then gauging the $U(N_f)$ flavor symmetry on both sides. While this involved careful consideration of the BF terms when $N_f=1$~\cite{Karch:2016sxi,Seiberg:2016gmd}, the case with non-Abelian global symmetries is actually simpler, since the $SU(N_f)$ subgroup of the flavor symmetry is manifest on both sides of the seed. It acts by simply rotating the matter fields.

This duality can be naively viewed as ``dualizing the first node'' of Theory A. That is, by performing the duality \eqref{base1} with $N_f= k_2$ flavors on the first of the two gauge group factors. As we will see, the shift in the CS level of the second gauge group factor is needed to reproduce the same topological field theory in the infrared after mass deformations on both sides.

Both sides of the new duality have a single conserved charge, which can be understood to be a monopole number. Naively the bosonic theory has a $U(1)\times U(1)$ monopole symmetry, arising from the $U(1)\times U(1)$ subgroup of the $U(k_1)\times U(k_2)$ gauge group, but only one linear combination acts faithfully, as in ABJM theory~\cite{Aharony:2008ug}.

We have so far glossed over an important point. In theory A the scalar field is bifundamental and so couples directly to both abelian factors of the $U(k_1)\times U(k_2)$ gauge theory. However, in the original base theory, $U(k_1)_{N_1}$ Chern-Simons theory coupled to $k_2$ scalars, the background field which couples to $U(1)$ global symmetry does so not through a coupling to the scalars, but through a BF term. So how can gauging the $U(k_2)$ flavor symmetry lead to theory A? Relatedly, is there a mixed Chern-Simons term between the abelian factors of theory A? To answer these questions, let us carefully show how this duality follows from the seed duality~\ref{base1}. We begin with
\begin{equation*}
U(k_1)_{N_1} \mbox{ with } k_2 \mbox{ WF scalars}\quad \leftrightarrow \quad SU(N_1)_{-k_1+\frac{k_2}{2}} \mbox{ with } k_2 \mbox{ Dirac fermions}
\end{equation*}
The scalar theory has a manifest $SU(k_2)\times U(1)$ global symmetry, where the $SU(k_2)$ rotates the scalars and the $U(1)$ is monopole number. The fermion theory has a manifest $U(k_2)=(SU(k_2)\times U(1))/\mathbb{Z}_{k_2}$ global symmetry. We identify the monopole current of the scalar theory with the baryon current of the fermion theory. Coupling the $SU(k_2)$ global symmetry to a background gauge field $A_{\mu}$ and the $U(1)$ to a background gauge field $\tilde{A}_{\mu}$, the Lagrangians for the scalar and fermion theories are~\cite{Hsin:2016blu}
\begin{align}
\begin{split}
\label{E:basicLagrangians}
\mathcal{L}_{\rm scalar} & = \frac{N_1}{4\pi}\text{tr}\left( ada+\frac{2}{3}a^3\right) + \mathcal{L}_{\Phi} + \frac{N_2}{4\pi}\text{tr}\left( A dA+\frac{2}{3}A^3\right) - \frac{N_1}{2\pi}\text{tr}(a)d\tilde{A} + \frac{N_1k_1+N_2k_2}{4\pi}\tilde{A}d\tilde{A}\,,
\\
\mathcal{L}_{\rm fermion} & = \frac{-k_1+k_2}{4\pi}\text{tr}\left( bdb+\frac{2}{3}b^3\right) + \mathcal{L}_{\Psi} + \frac{N_1+N_2}{4\pi}\text{tr}\left( A dA+\frac{2}{3}A^3\right) + \frac{k_2(N_1+N_2)}{4\pi}\tilde{A}d\tilde{A}\,,
\end{split}
\end{align}
where $\mathcal{L}_{\Phi}$ is the kinetic term for the scalars along with a critical potential and $\mathcal{L}_{\Psi}$ is the kinetic term for the fermions. Here $a_{\mu}$ is a $U(k_1)$ gauge field and $b_{\mu}$ an $SU(N_1)$ gauge field, and the various gauge fields couple to the scalars and fermions through covariant derivatives
\begin{align}
\begin{split}
D_{\mu}\Phi & = \left( \partial_{\mu} + i (a_{\mu} \mathbbm{1}_f + A_{\mu}\mathbbm{1}_c)\right)\Phi\,,
\\
D_{\mu}\Psi & = \left(\partial_{\mu} + i (b_{\mu} \mathbbm{1}_f + A_{\mu}\mathbbm{1}_c + \tilde{A}_{\mu}\mathbbm{1})\right)\Psi\,,
\end{split}
\end{align}
with $\mathbbm{1}_f$ the identity acting on flavor indices, $\mathbbm{1}_c$ the identity acting on color indices, and $\mathbbm{1}$ the identity. The flavor counterterms in both Lagrangians have been matched. In the scalar theory, decomposing the $U(k_1)$ gauge field into an $SU(k_1)$ part $a'$ and abelian part $\tilde{a}$,
\beq
a_{\mu} = a'_{\mu} + \tilde{a}_{\mu}\mathbbm{1}\,,
\eeq
we are free to shift the abelian part of $a$ by $\tilde{A}$,
\beq
\tilde{a} \to \tilde{a} + \tilde{A}\,.
\eeq
Doing so, the scalar Lagrangian becomes
\begin{align}
\begin{split}
\label{E:newScalarL}
\mathcal{L}_{\rm scalar} &= \frac{N_1}{4\pi} \text{tr}\left( ada+\frac{2}{3}a^3\right) + \mathcal{L}_{\Phi}  + \frac{N_2}{4\pi}\text{tr}\left( AdA + \frac{2}{3}A^3\right) + \frac{N_2k_2}{4\pi}\tilde{A}d\tilde{A}\,,
\\
D_{\mu}\Phi & = \left( \partial_{\mu} +i (a_{\mu} \mathbbm{1}_f + A_{\mu} \mathbbm{1}_c + \tilde{A}_{\mu} \mathbbm{1})\right)\Phi\,,
\end{split}
\end{align}
Now, the last two topological terms in the fermion Lagrangian~\eqref{E:basicLagrangians} as well as the scalar Lagrangian~\eqref{E:newScalarL} are $U(k_2)$ Chern-Simons terms for a $U(k_2)$ gauge field
\beq
C_{\mu} \equiv A_{\mu} + \tilde{A}_{\mu}\mathbbm{1}\,,
\eeq
i.e.
\begin{align}
\begin{split}
\mathcal{L}_{\rm scalar} & = \frac{N_1}{4\pi}\text{tr}\left( ada+\frac{2}{3}a^3\right) + \mathcal{L}_{\Phi} + \frac{N_2}{4\pi}\text{tr}\left( CdC+\frac{2}{3}C^3\right)\,,
\\
\mathcal{L}_{\rm fermion} & = \frac{-k_1+k_2}{4\pi}\text{tr}\left( bdb+\frac{2}{3}b^3\right) + \mathcal{L}_{\Psi} + \frac{N_1+N_2}{4\pi}\text{tr}\left( CdC+\frac{2}{3}C^3\right)\,,
\end{split}
\end{align}
with $\Phi$ bifundamental under $U(k_1)\times U(k_2)$ and $\Psi$ under $SU(k_1)\times U(k_2)$. Gauging the $U(k_2)$ global symmetry on both sides leads to theories A and B, and the base duality equates them. Observe that there is no mixed abelian Chern-Simons term in theory A.

\subsection{Deformations}

Our conjecture \eqref{dualone} postulates an equivalence between two interacting CFTs and so is difficult to verify. One simple test is to add relevant mass deformations to the CFT. In 2+1 dimensional CS-matter theories such massive deformations often result in non-trivial TFTs. Furthermore, the particular TFT depends crucially on the sign of the mass deformation, so that the original CFT describes a phase transition between two distinct TFTs. Upon deformation we ought to reproduce the same TFTs on both sides of the putative duality. This is a stringent test of any proposal. 

The seed dualities~\eqref{base1} and~\eqref{base2} already pass this test. Not only are the TFTs level/rank dual, but the flavor Chern-Simons terms in the massive phases also match on the nose. Since our duality comes from gauging flavor symmetries on both sides of the seed duality, it should also pass this test.

Let us see how this works in detail, starting with theory A. The relevant deformation of interest is a scalar mass $m$. For a positive mass squared, the scalar simply decouples (phase A1). For a negative mass-squared, the scalar field condenses. We assume that the preferred scalar vacuum expectation value (vev) is the one that maximally Higgses the gauge group. For a product $U(k)_{N_1} \times U(k)_{N_2}$ gauge group a generic bifundamental scalar vev breaks the gauge group to the diagonal $U(k)_{N_1 + N_2}$ subgroup where the CS terms simply add. When the gauge groups have unequal rank the scalar vev breaks the common $U(k_2)$ factor to its diagonal subgroup. Once again, the CS terms simply add. The remaining $U(k_1-k_2)_{N_1}$ factor of the larger $U(k_1)_{N_1}$ gauge group is untouched. The net result is that we can view our CFT in theory A as describing the phase transition between the following two TFTs:
\begin{align}
\begin{split}
\label{E:theoryATFTs}
(A1)  \quad \quad  m^2 > 0: &\quad   U(k_1)_{N_1} \times U(k_2)_{N_2}  \\
(A2)  \quad \quad m^2 < 0: &\quad   U(k_1-k_2)_{N_1} \times U(k_2)_{N_1+N_2}
\end{split}
\end{align}
In the base duality \eqref{base1} it is well understood how to map the scalar mass-squared to the dual theory. A scalar mass deformation simply maps to a fermion mass $M$. The sign is reversed compared to theory A: a positive scalar mass-squared leads to a negative fermion mass and vice versa. In either case, we can simply integrate out the massive fermions, which crucially leads to a shift $\sign(M)\,N_f/2$ in the Chern-Simons terms. Correspondingly, we obtain the following phases in theory B:
\begin{eqnarray}
\nonumber
(B1)  \quad \quad  M < 0: &\quad &  SU(N_1)_{-k_1} \times U(k_2)_{N_2}  \\
\nonumber
(B2)  \quad \quad M > 0: &\quad &  SU(N_1)_{k_2-k_1} \times U(k_2)_{N_1+N_2}
\end{eqnarray}
Level-rank duality equates $U(k)_N$ and $SU(N)_{-k}$, so we see that (A1) and (B1) as well as (A2) and (B2) describe the same TFTs respectively. This gives significant evidence to the conjecture that the CFTs of theory A and B describe the same phase transition between these distinct topological phases.

\subsection{The special case $k_1=k_2$ and $N_1=-N_2$}

As we have seen, the flavor bound $N_f \leq k$ of the base pair \eqref{base1} implies that the duality \eqref{dualone} of the previous subsection is valid for $k_1 \leq k_2$. As we have noted, this is not a significant restriction since we can always, without loss of generality, chose the gauge group factor with the bigger CS level to be the first one. The limiting case of $k_1=k_2$, together with $N_1=-N_2$ is special and deserves a separate discussion. Phase (A1) is still a non-trivial TFT, but phase 2 is different. We are left with a single $U(k_2)_0$ gauge theory without CS term. The non-Abelian factor of 2+1 dimensional pure glue is believed to confine and does not contribute any non-trivial physics in the IR. The $U(1)_0$ factor however contributes a massless photon. So in this special case, phase (A2) is in fact no longer a TFT but describes a free Coulomb phase of a single massless scalar with a shift symmetry (the dual photon).

In theory B phase (B2) has two apparently non-trivial gauge group factors, but both of them have level 0 and hence do not contribute to the IR. As in theory A, only the $U(1)$ factor of the second gauge group contributes to the low energy physics and, while no longer describing a TFT, phase (B2) is still identical to phase (A2).

\section{Quivers}

\subsection{Linear quivers}

The same strategy can be employed to generate more interesting examples involving products of gauge group factors coupled by bifundamental matter: quiver gauge theories. A simple example involving three gauge groups is the following ``linear quiver" theory:

\vskip10pt
\begin{center}
\begin{tabular}{|c|c|c|c|}
\hline
\multicolumn{4}{|c|}{\bf Theory A} \\
\hline
\hline
&$U(k_1)_{N_1}$ & $U(k_2)_{N_2}$ & $U(k_3)_{N_3}$ \\
\hline
$X_1$ & $\Box$ & $\Box$ & 1 \\
$X_2$ & 1 & $\Box$ & $\Box$ \\
\hline
\end{tabular}
\vskip10pt
\quad {\huge $\updownarrow$} \quad
\vskip10pt
\begin{tabular}{|c|c|c|c|}
\hline
\multicolumn{4}{|c|}{\bf Theory B} \\
\hline
\hline
&$U(k_1)_{N_1 + \frac{N_2}{2}}$&$SU(N_2)_{-k_2 + \frac{k_1 + k_3}{2}}$ & $U(k_3)_{N_3+ \frac{N_2}{2}}$ \\
\hline
$\psi_1$ &$\Box$ & $\Box$ & 1 \\
$\psi_2$ & 1&$\Box$ & $\Box$ \\
\hline
\end{tabular}
\end{center}

In this example we performed the base duality \eqref{base1} on the middle node or, to be more precise, we started with the base duality between $SU(N_2)$ theory with $N_f=k_1+k_3$ fermions and $U(N_2)$ theory with $N_f$ scalars, and gauged a $U(k_1) \times U(k_3)$ subgroup of the global $U(k_1+k_3)$ flavor group. Correspondingly, the duality should hold as long as $k_2\geq k_1+k_3$. One new subtlety here is that there are extra deformations on the scalar side. In addition to the two quartics $|X_1|^4$ and $|X_2|^4$ that we obviously need to add since the duality always applies to WF scalars, we now also have to tune the two deformations $|X_1|^2 |X_2|^2$ and $|X_1X_2^*|^2$. This extra information is needed to completely specify the bosonic CFT. Our claim is that with appropriate tuning of the quartic couplings the system flows to a CFT that is equivalent to the fermionic theory which is uniquely specified by the gauge groups, CS levels, and matter content.

As in the previous example, we can test our duality by adding relevant deformations to drive the CFT into topological phases. Since this time we have two different fields we can give mass to, the CFT should be thought of as a multi-critical point that can connect to four different topological phases depending on the various sign choices. In the scalar theory we again can simply drop scalars with positive mass-squared, whereas bifundamentals with negative mass-squared condense and break neighboring gauge groups to the diagonal subgroup as far as possible, leaving a part of the original larger rank gauge group unbroken. On the fermion side we simply integrate out the fermions, shifting the CS levels of the gauge groups under which they are charged depending on the sign of the mass. The various TFT phases in theory A are:
\begin{eqnarray}
\nonumber
(A1)  \quad \quad  m_1^2 > 0, \quad m_2^2>0: &\quad &  U(k_1)_{N_1} \times U(k_2)_{N_2} \times U(k_3)_{N_3}  \\
\nonumber
(A2)  \quad \quad m_1^2>0, \quad m_2^2 < 0: &\quad &  U(k_1)_{N_1} \times U(k_2-k_3)_{N_2} \times U(k_3)_{N_2+N_3} \\
\nonumber
(A3)  \quad \quad  m_1^2 < 0, \quad m_2^2>0: &\quad &  U(k_1)_{N_1+N_2} \times U(k_2-k_1)_{N_2} \times U(k_3)_{N_3}  \\
\nonumber
(A4)  \quad \quad m_1^2<0, \quad m_2^2 < 0: &\quad &  U(k_1)_{N_1+N_2} \times U(k_2-k_3-k_1)_{N_2} \times U(k_3)_{N_2+N_3}
\end{eqnarray}
whereas the corresponding phases of theory B are
\begin{eqnarray}
\nonumber
(B1)  \quad \quad  M_1 < 0, \quad M_2<0: &\quad &  U(k_1)_{N_1} \times SU(N_2)_{-k_2} \times U(k_3)_{N_3}  \\
\nonumber
(B2)  \quad \quad M_1<0, \quad M_2 > 0: &\quad &  U(k_1)_{N_1} \times SU(N_2)_{k_3-k_2} \times U(k_3)_{N_2+N_3} \\
\nonumber
(B3)  \quad \quad  M_1  > 0, \quad M_2<0: &\quad &  U(k_1)_{N_1+N_2} \times SU(N_2)_{k_1-k_2} \times U(k_3)_{N_3}  \\
\nonumber
(B4)  \quad \quad M_1>0, \quad M_2 > 0: &\quad &  U(k_1)_{N_1+N_2} \times SU(N_2)_{k_3+k_1-k_2} \times U(k_3)_{N_2+N_3}
\end{eqnarray}

Once again, these topological phases in theories A and B perfectly agree with each other upon using level/rank duality. This gives evidence to our conjecture that the multi-critical point at which all four phases meet agrees between the fermionic and bosonic description.

\subsection{Circular quivers}

Clearly our method can be used to generate infinitely many new conjectures by dualizing nodes in a quiver diagram, one at a time. Let us end this Section with an example of a circular quiver with four nodes, valid for $2n \leq k_1$:
\vskip10pt
\begin{center}
\begin{tabular}{|c|c|c|c|c|}
\hline
\multicolumn{5}{|c|}{\bf Theory A} \\
\hline
\hline
&$U(k_1)_{N}$ & $U(n)_{k_2}$ & $U(k_1)_{N}$ & $U(n)_{k_2}$ \\
\hline
$X_1$ & $\Box$ & $\Box$  & 1     &1 \\
$X_2$ & 1     & $\Box$  & $\Box$ &1\\
$X_3$ & 1     & 1      & $\Box$ & $\Box$ \\
$X_4$ & $\Box$ & 1      & 1     & $\Box$\\
\hline
\end{tabular}
\vskip10pt
\quad {\huge $\updownarrow$} \quad
\vskip10pt
\begin{tabular}{|c|c|c|c|c|}
\hline
\multicolumn{5}{|c|}{\bf Theory B} \\
\hline
\hline
&$SU(N)_{-k_1+n}$ & $U(n)_{k_2+N}$ & $SU(N)_{-k_1+n}$ & $U(n)_{k_2+N}$ \\
\hline
$\psi_1$ & $\Box$ & $\Box$  & 1     &1 \\
$\psi_2$ & 1     & $\Box$  & $\Box$ &1\\
$\psi_3$ & 1     & 1      & $\Box$ & $\Box$ \\
$\psi_4$ & $\Box$ & 1      & 1     & $\Box$\\
\hline
\end{tabular}
\end{center}

This duality can be derived by starting with two decoupled $SU(N)_{-k+n}$ factors with $2n$ flavors of fermions each, dual to two decoupled $U(k)_N$ factors with $2n$ scalars each. We then stitch the two factors together by gauging a common $U(n) \times U(n)$ subgroup of the two $U(2n)$ flavor symmetries. Equivalently, we start with the circular quiver in theory A, and then dualize the first and third nodes. Understanding the full phase structure of this theory is difficult as we need to be able to identify breaking patterns when the various scalar fields get vevs. The simple topological phase in which all scalars get a positive mass-squared (and hence all fermions get negative mass) clearly agrees in both theories:
$$
(A1)=(B1): \quad \quad U(k_1)_{N} \times U(n)_{k_2} \times U(k_1)_{N} \times U(n)_{k_2}\,.
$$
All other phases can easily be determined from the fermionic side. To reproduce, say, the phase in which all fermions get positive masses (and hence all scalars condense) we need to assume that the correct breaking pattern in the bosonic theory is as follows:
$$
[U(k_1)_{N}]^2 \times [U(n)_{k_2}]^2  \rightarrow [U(k_1-2n)_N]^2 \times [ U(n)_{k_1+2N} ]^2\,.
$$
That is, we break both $U(k_1)$ gauge groups down to a diagonal $U(n)^2$ factor which then is also identified with the remaining $U(n)^2$. As always, the CS levels add up to $2N+k_1$. The remaining $U(k_1-2n)$ subgroups remain unbroken.

\section{Bose/Bose and Fermi/Fermi dualities}

All dualities we presented in the previous Section related scalars to fermions. In the spirit of dualizing one node at a time we can also derive Bose/Bose and Fermi/Fermi dualities by starting with a parent theory, dualizing the first node as above and then going back to the parent and dualizing the second node. More precisely, we can obtain one and the same fermionic parent theory P by starting with two different bosonic theories and gauging the global symmetry. In one case it is the first node of theory P that appears as the gauged global symmetry, in the second case it is the second node. This way the fermionic theory P has separate bosonic duals A and B, implying a Bose/Bose duality between A and B. Similarly one can expect Fermi/Fermi dualities by starting with a bosonic parent P. As we will see below, the Fermi/Fermi case is much more constrained by the flavor bounds.

\subsection{Bose/Bose duality}
\label{sec:bosebose}

Let us just present one example in this spirit. Theories A and B of \eqref{dualone} have another partner with scalar matter, theory C, which is given by
\vskip10pt
\begin{center}
\begin{tabular}{|c|c|c|}
\hline
\multicolumn{3}{|c|}{\bf Theory C} \\
\hline
\hline
&$SU(N_1)_{-k_1+k_2}$ & $SU(N_1+N_2)_{-k_2}$ \\
\hline
$Y$ & $\Box$ & $\Box$ \\
\hline
\end{tabular}
\end{center}
\vskip10pt
We can think of theory C arising from applying the duality \eqref{base2} on the second factor in theory B with $k=k_2$, $N=N_1+N_2$ and $N_f=N_1$ (and hence we need to require that $N_1 \leq k_2$). The CFT in C connects topological phases
\begin{align}
\begin{split}
\label{E:theoryCTFTs}
(C1)  \quad \quad  m_Y^2 < 0: &\qquad  SU(N_1)_{-k_1} \times SU(N_2)_{-k_2}  \\
(C2)  \quad \quad m_Y^2 > 0: &\qquad  SU(N_1)_{k_2-k_1} \times SU(N_1+N_2)_{-k_2}
\end{split}
\end{align}
which are just the level/rank duals of the TFTs in phases (A1) and (A2) of theory A,~\eqref{E:theoryATFTs}, as expected.

In the special case $k_1=k_2=k$, theory C becomes $SU(N)_0\times SU(N_1+N_2)_{-k}$ Chern-Simons theory with a bifundamental scalar. Under the expectation that the $SU(N)_0$ factor confines, theory C ought to flow to the low-energy CFT
\beq
\textbf{Theory C: (when $k_1=k_2=k$)} \longrightarrow SU(N_1+N_2)_{-k} \quad +\quad  \text{adjoint}\,.
\eeq
The TFT in the Higgs phase depends on how the adjoint scalar condenses. Assuming that the dynamics prefers to break $SU(N_1+N_2)\to SU(N_1)\times SU(N_2)$ we find that the massive phases are exactly those above in~\eqref{E:theoryCTFTs} when $k_1=k_2=k$:
\begin{eqnarray}
\nonumber
(C1) \quad \quad m^2_Y<0: & \quad & SU(N_1)_{-k} \times SU(N_2)_{-k}
\\
\nonumber
(C2) \quad \quad m^2_Y >0: & \quad & SU(N_1+N_2)_{-k}
\end{eqnarray}

\subsection{Fermi/Fermi duality}
\label{SS:fermi}

Naively one might think that the derivation of a Fermi/Fermi duality starting with a bosonic parent would proceed just as in the Bose/Bose case above. This is not quite true on account of the flavor bound. Note that the flavor bound $N_f \leq k$ of the base pairs \eqref{base1} and \eqref{base2} states that the number of flavors must to be less than the level of the fermionic theory, equivalently less than the rank in the bosonic theory. In the product gauge case, the rank of one gauge group factor gives the number of flavors in the second gauge group factor. The level on the other hand can be freely dialed without changing the flavor content of either gauge group. So if we start with a fermionic parent, we can always make sure that the flavor bound is obeyed by both gauge group factors by adjusting the level accordingly. For example, in the Bose/Bose duality of the last Subsection, we had to choose $N_1 \leq k_2$  in addition to $k_2 \leq k_1$. In contrast, if we start with a bosonic parent theory with $U(k_1) \times U(k_2)$ gauge group (or another theory with either or both factors replaced replaced by an $SU$ group), we need $k_2 \leq k_1$ and $k_1 \leq k_2$ to be true simultaneously if we want to make sure that there are two distinct Fermi duals. So we should only expect to be able to derive a Fermi/Fermi duality for the special case of $k=k_1=k_2$. So instead of a four-parameter family ($N_1$, $N_2$, $k_1$, $k_2$) of Bose/Bose dualities, we can only construct a three-parameter family of Fermi/Fermi duals.

Starting from a scalar parent theory P that is just theory A from \eqref{dualone} with $k_1=k_2=k$,
\vskip10pt
\begin{center}
\begin{tabular}{|c|c|c|}
\hline
\multicolumn{3}{|c|}{\bf Theory P} \\
\hline
\hline
&$U(k)_{N_1}$ & $U(k)_{N_2}$ \\
\hline
$X$ & $\Box$ & $\Box$ \\
\hline
\end{tabular}
\end{center}
we can derive the following Fermi/Fermi duality by dualizing on the first and second node respectively
\vskip10pt
\begin{center}\hfill
\begin{minipage}[h]{0.35\hsize} \centering
\begin{tabular}{|c|c|c|}
\hline
\multicolumn{3}{|c|}{\bf Theory A} \\
\hline
\hline
&$SU(N_1)_{-\frac{k}{2}}$ & $U(k)_{N_2 + \frac{N_1}{2}}$ \\
\hline
$\psi_A$ & $\Box$ & $\Box$ \\
\hline
\end{tabular}
\end{minipage}
\quad {\huge $\leftrightarrow$} \quad
\begin{minipage}[h]{0.32\hsize}\centering
\begin{tabular}{|c|c|c|}
\hline
\multicolumn{3}{|c|}{\bf Theory B} \\
\hline
\hline
&$U(k)_{N_1+\frac{N_2}{2}}$ & $SU(N_2)_{-\frac{k}{2}}$ \\
\hline
$\psi_B$ & $\Box$ & $\Box$ \\
\hline
\end{tabular}
\end{minipage}
\hfill
\begin{minipage}[h]{0.1\hsize} \beq \label{fermifermi} \eeq \end{minipage}
\end{center}
\vskip10pt
As in all our dualities, we can easily confirm that these putative dual CFTs sit at the critical point between identical topological phases which we get by mass deformations:
\begin{eqnarray}
\nonumber
(A1)  \quad \quad  m_A > 0: &\quad &  SU(N_1)_{0} \times U(k)_{N_1+N_2}  \\
\nonumber
(A2)  \quad \quad m_A < 0: &\quad &  SU(N_1)_{-k} \times U(k)_{N_2}\\
\nonumber
(B1)  \quad \quad  m_B > 0: &\quad &  U(k)_{N_1 + N_2} \times SU(N_2)_{0}  \\
\nonumber
(B2)  \quad \quad m_B < 0: &\quad &  U(k)_{N_1} \times SU(N_2)_{-k}
\end{eqnarray}
Since $SU(N)_0$ is a trivial confining sector irrespective of $N$, we see that (A1) and (B1) are tirivally identical phases, whereas (A2) and (B2) are level/rank duals.

\section{Conclusion and Discussion}

We presented a new method to generate infinitely many new conjectured non-supersymmetric dualities in 2+1 dimensions. We performed some very basic consistency checks and our proposed dualities passed with flying colors. 

There are clearly many more dualities one can obtain this way. We just presented a few simple concrete examples of products involving unitary and special unitary groups. One can apply the same techniques to generate conjectures for quivers of almost arbitrary topology by dualizing a single node at a time (as long as it obeys the relevant flavor constraints). Using the dualities presented in \cite{Aharony:2016jvv,Metlitski:2016dht} we can also handle products with $SO(N)_k$ or $USp(2N)_k$ factors.

We conclude with something that may be even more interesting. It seems we can get dualities for single gauge groups with matter in representations other than the fundamental. In principle we can derive these from our product group dualities via orbifold identifications. Let us briefly sketch one such construction. It will be clear that while the details need to be better understood, this simple example gives us already new conjectured dualities that are consistent with level/rank duality.

Let us start with the base dualities of Subsections~\ref{basicduality} and~\ref{sec:bosebose} for $k_1=k_2 \equiv k$ and $N_1=N_2\equiv N$. These combine into a triality:\footnote{When $k_1=k_2$ there is also a Fermi/Fermi duality as we discussed in Subsection~\ref{SS:fermi} and so for general $N_1, N_2$, there is actually a quadrality between two bosonic theories and two fermionic ones. However both fermionic theories are identical when $N_1=N_2$.}
\vskip10pt
\begin{center}\hfill \hspace{.68in}
\begin{minipage}[h]{0.35\hsize} \centering
\begin{tabular}{|c|c|c|}
\hline
\multicolumn{3}{|c|}{\bf Theory A} \\
\hline
\hline
&$U(k)_{N}$ & $U(k)_{N}$ \\
\hline
$X_A$ & $\Box$ & $\Box$ \\
\hline
\end{tabular}
\end{minipage}
{\huge $\leftrightarrow$} \quad
\begin{minipage}[h]{0.32\hsize}\centering
\begin{tabular}{|c|c|c|}
\hline
\multicolumn{3}{|c|}{\bf Theory B} \\
\hline
\hline
&$SU(N)_{-k/2}$ & $U(k)_{3N/2}$ \\
\hline
$\psi_B$ & $\Box$ & $\Box$ \\
\hline
\end{tabular}
\end{minipage}
\hfill\begin{minipage}[h]{0.1\hsize} \beq \label{orbiparent} \eeq \end{minipage}
\vskip10pt
\quad {\huge $\updownarrow$} \quad
\vskip 10pt
\begin{minipage}[h]{0.32\hsize}\centering
\hfill
\begin{tabular}{|c|c|c|}
\hline
\multicolumn{3}{|c|}{\bf Theory C} \\
\hline
\hline
&$SU(N)_{0}$ & $SU(2 N)_{-k}$ \\
\hline
$X_C$ & $\Box$ & $\Box$ \\
\hline
\end{tabular}
\end{minipage}
\hfill
\end{center}
The fields $X_A$ and $X_C$ are WF scalars and $\psi_B$ is a Dirac fermion. As we discussed in Subsection~\ref{sec:bosebose}, we expect the $SU(N)_0$ gauge group factor in theory C to confine, so that its low-energy limit is
\begin{equation*}
\textbf{Theory C:} \quad SU(2N)_{-k} \quad +\quad  \text{adjoint}\,.
\end{equation*}
Obviously theory A has a $\mathbb{Z}_2$ symmetry $E$ exchanging the two $U(k)$ factors. If we mod out by this we get
\beq
\nonumber
\textbf{Theory A'}: \quad U(k)_N  \quad + \quad  \Box \hspace{-1.2pt} \Box\,.
\eeq
Assuming the triality is correct, $E$ must also be a symmetry of theories B and C. However, $E$ is not manifest there, and so we conclude that it is an emergent symmetry. We would like to orbifold theories B and C by it, but since $E$ is not manifest some guesswork is required to figure out the effect of the orbifold. We will come to theory B soon, but there is a natural conjecture for the orbifold in theory $C$. It ought to give:
\beq
\textbf{Theory C'}: \quad SO(2N)_{-k} \quad + \quad  \mbox{adjoint}\,.
\eeq
So we propose an equivalence between theories A' and C'.

We can put this conjecture to the test using our standard strategy of working out the mass deformations. The CFTs of theory A' and C' mediate phase transitions between the following topological phases respectively
\begin{eqnarray}
\nonumber
(a1)  \quad \quad  M^2_A > 0: &\quad &    U(k)_N\\
\nonumber
(a2)  \quad \quad M^2_A < 0: &\quad & SO(k)_{2N}\\
\nonumber
(c1)  \quad \quad  M^2_C >0: &\quad & SO(2N)_{-k}  \\
\nonumber
(c2)  \quad \quad M^2_C < 0: &\quad & SU(N)_{-k}
\end{eqnarray}
Reassuringly (a1) and (c2) and well as (a2) and (c1) are the same TFTs by level/rank duality. So at the level of rigor at which we were able to test the dualities proposed in this work, the conjecture based on the orbifold construction also passes this simple check.

There should also be an orbifolded version of the fermionic theory B, but we do not know how to obtain it. The assumption that such a fermionic theory exists, however, is quite constraining, and leads us to conjecture that:
\vskip10pt
\begin{center}
\begin{tabular}{|c|c|c|}
\hline
\multicolumn{3}{|c|}{\bf Theory B'} \\
\hline
\hline
&$SU(N)_{-k/2}$ & $SO(k)_{N}$ \\
\hline
$\psi_B$ & $\Box$ & $\Box$ \\
\hline
\end{tabular}
\end{center}
where $\psi_B$ is a Dirac fermion, is dual to theories A' and C'. Let us first check that the global symmetries and massive phases match, and then we can explain how we came to theory B'. First, all three theories have a $U(1)$ global symmetry: in theory A' there is a conserved monopole number, while theories B' and C' have a conserved baryon number (except when $k=2$ or $N=2$, in which case theory B' and C' respectively has a monopole number). Second, the massive phases are:\footnote{We are using here that integrating out a fundamental massive Dirac fermion shifts the level of an $SO(k)$ Chern-Simons term by $\pm \text{sgn}(m)$.}
\begin{eqnarray}
\nonumber
(b1) \quad \quad m_B>0: & \quad & SU(N)_0\times SO(k)_{2N}  \\
\nonumber
(b2) \quad \quad m_B<0: & \quad & SU(N)_{-k} \times SO(k)_0
\end{eqnarray}
As long as $k\neq 2$, we see that in both phases one of the gauge groups is at level 0 and so confines, leading to the TFT $SO(k)_{2N}$ in phase (b1) and $SU(N)_{-k}$ in phase (b2). These agree with phases (a2)/(c1) and (a1)/(c2) respectively by level/rank duality. We do not know what to expect when $k=2$.

The logic that led us to theory B' is simple. Integrating out a massive fermion of a Chern-Simons matter theory shifts the level of a Chern-Simons term, but, barring a transition to a color superconducting phase, doesn't alter the gauge group. This, combined with the requirement of matching the massive bosonic phases, is a stringent constraint. Clearly a Chern-Simons theory with a single gauge group cannot do the job, as one phase must be a TFT based on a $U/SU$ theory and the other based on an $SO$ theory. The only type of quiver gauge theory that can realize this with only two phases is a two-node quiver, with levels tuned so that the level of one gauge group vanishes for one sign of Dirac mass, and the level of the other vanishes for the other sign. This is exactly what happens in theory B', and it is easy to check that there is no other fermionic two-node quiver with the same massive phases. So theory B' is the unique fermionic Chern-Simons matter theory whose massive phases match those of theories A' and C'.

\section*{Acknowledgements}

We would like to thank Ofer Aharony for illuminating correspondence. The work of KJ and AK was supported in part by the US Department of Energy respectively under grant numbers DE-SC0013682 and DE-SC0011637.

\bibliographystyle{JHEP}
\bibliography{stringmirror}

\end{document}